\documentstyle[sprocl]{article}

\input{psfig}

\def\be{\begin{equation}}
\def\ee{\end{equation}}
\def\bea{\begin{eqnarray}}
\def\eea{\end{eqnarray}}

\def\GeV{\,{\rm GeV}}

\def\sec{\,{\rm sec}}
\def\Gyr{\,{\rm Gyr}}

\def\rcm{\,{\rm cm}}

\def\Mpc{\,{\rm Mpc}}

\def\eV{{\,\rm eV}}

\def\cmm2{{\,\rm cm^{-2}}}
\def\cm2{{\,{\rm cm}^2}}
\def\cmm3{{\,{\rm cm}^{-3}}}
\def\gcmm3{{\,{\rm g\,cm^{-3}}}}
\def\kms{\,{\rm km\,s^{-1}}}

\def\mpl{{m_{\rm Pl}}}

\def\la{\mathrel{\mathpalette\fun <}}

\def\fun#1#2{\lower3.6pt\vbox{\baselineskip0pt\lineskip.9pt
  \ialign{$\mathsurround=0pt#1\hfil##\hfil$\crcr#2\crcr\sim\crcr}}}

\begin{document}

\title{COSMOLOGY UPDATE 1998}

\author{Michael S. TURNER}

\address{Departments of Astronomy \& Astrophysics and of Physics\\
Enrico Fermi Institute, The University of Chicago\\
Chicago, IL~~60637-1433, USA\\
E-mail: mturner@oddjob.uchicago.edu}

\address{NASA/Fermilab Astrophysics Center\\
Fermi National Accelerator Laboratory, Box 500\\
Batavia, IL~~60510-0500, USA}


\maketitle\abstracts{For two decades the hot big-bang model
has been referred to as the standard cosmology -- and for good reason.
For just as long
cosmologists have known that there are fundamental questions
that are not addressed by the standard cosmology and point to
a grander theory.  The best candidate for that grander
theory is inflation + cold dark matter.
It holds that the Universe is flat, that slowly moving elementary
particles left over from the earliest moments provide
the cosmic infrastructure, and that the primeval density
inhomogeneities that seed
all large-scale structure arose from quantum fluctuations.
There is now prima facie evidence that supports
two basic tenets of this paradigm, and an avalanche of
high-quality cosmological observations will soon make this case
stronger or will break it.  Key questions remain to be
answered; foremost among them are:  identification and
detection of the cold dark
matter particles and elucidation of the mysterious
dark-energy component.
These are exciting times in cosmology!}

\section{1998, A Remarkable Year for Cosmology}

The birth of the hot big-bang model dates back to the work of Gamow
and his collaborators in the 1940s.  The emergence of the hot big-bang
cosmology began in 1964 with the discovery of
the microwave background radiation.  By the 1970s,
the black-body character of the microwave background radiation
had been established and the success of big-bang nucleosynthesis
demonstrated, and the hot big-bang was being referred to as the standard
cosmology.  Today, it is universally accepted and provides an accounting
of the Universe from a fraction of a second after the beginning,
when the Universe was a hot, smooth soup of quarks and leptons to the
present, some $14\Gyr$ later.  Together with the standard model
of particle physics and ideas about the unification
of the forces, it provides a firm foundation for speculations
about the earliest moments of creation.

The standard cosmology rests upon
three strong observational pillars:  the expansion of the Universe; the cosmic
microwave background radiation (CBR); and the abundance pattern
of the light elements, D, $^3$He, $^4$He, and $^7$Li,
produced seconds after the bang (see e.g., Peebles et al, 1991;
or Turner \& Tyson, 1999).
In its success, it has raised new, more profound questions:
the origin of the matter/antimatter asymmetry, the origin of the smoothness
and flatness of the Universe, the nature and origin of the
primeval density inhomogeneities that seeded all the structure
in the Universe, the quantity and composition of the dark matter
that holds the Universe together, and the nature of the big-bang
event itself.  This has motivated the search for a more
expansive cosmological theory.

In the 1980s, born of the inner space/outer space connection,
a new paradigm emerged, one deeply rooted in fundamental
physics with the potential to extend our understanding of the Universe
back to $10^{-32}\sec$ and to address the fundamental questions
posed, but not addressed by the hot big-bang model.
That paradigm, known as inflation
+ cold dark matter, holds that most of the dark matter consists of
slowly moving elementary particles (cold dark matter),
that the Universe is flat and
that the density perturbations that seeded all the structure seen
today arose from quantum-mechanical fluctuations on scales of $10^{-23}
\rcm$ or smaller.  It took awhile for the observers and experimentalists
to take this theory seriously enough to try to disprove it, and in the 1990s
it began to be tested in a serious way.

\begin{figure}
\centerline{\psfig{figure=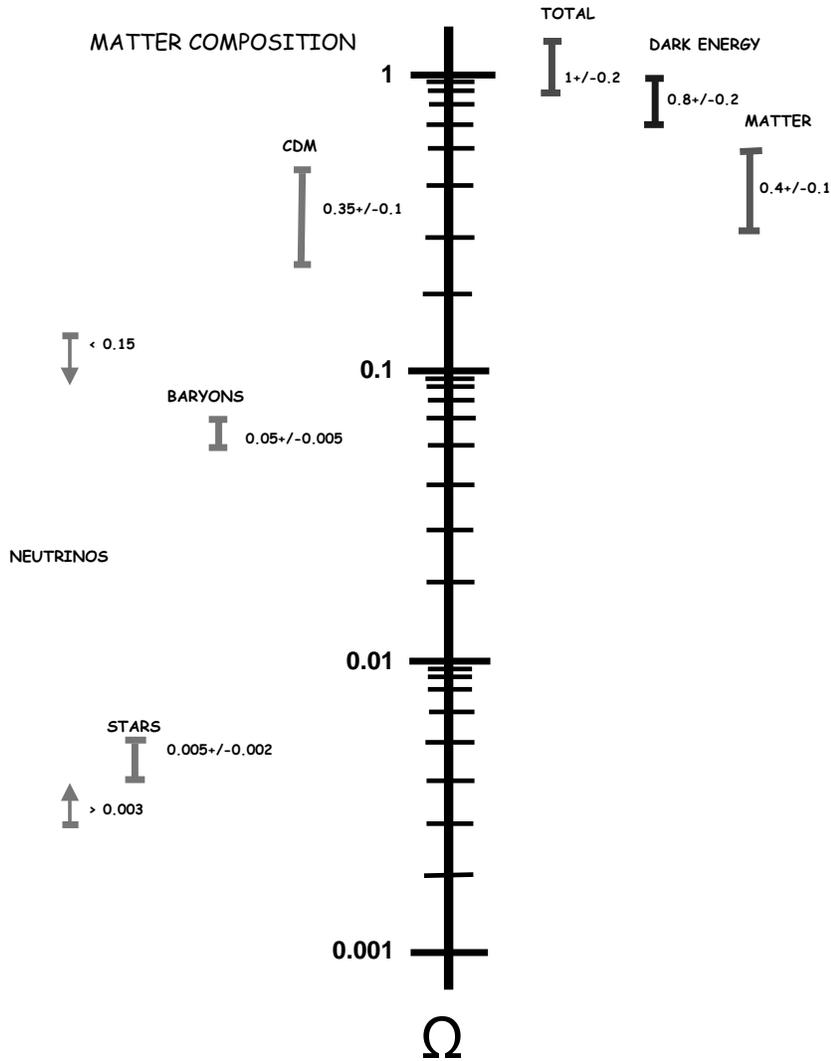,width=4.3in}}
\caption{Summary of matter/energy in the Universe.
The right side refers to an overall accounting of matter
and energy; the left refers to the composition of the matter
component.  The upper limit to mass density contributed
by neutrinos is based upon the failure of the hot dark
matter model of structure formation (Dodelson et al, 1996;
White, Frenk \& Davis, 1983)
and the lower limit follows from the
evidence for neutrino oscillations (Fukuda et al, 1998).
Here $H_0$ is taken to be $65\kms\Mpc^{-1}$.
}
\label{fig:omega_sum}
\end{figure}

This could prove to be a watershed year in cosmology, as important
as 1964, when the CBR was discovered.  The crucial
new data include a precision measurement of the
density of ordinary matter and of the total amount
of matter, both derived from a measurement of
the primeval deuterium abundance and the theory of BBN;
fine-scale (down to $0.3^\circ$) measurements of
the anisotropy of the CBR; and a measurement of the deceleration
of the Universe based upon distance measurements of
type Ia supernovae out to redshift of close to unity.

Together, these measurements, which are
harbingers for the precision era of cosmology that is coming,
provide the first plausible, complete accounting of the matter/energy
density in the Universe and evidence that the primeval density perturbations
arose from quantum fluctuations during inflation.
In addition, there exists a large body of cosmological data -- from
measurements of large-scale structure to the evolution of galaxies
and clusters -- that supports the cold dark matter theory of structure
formation.

\begin{figure}
\centerline{\psfig{figure=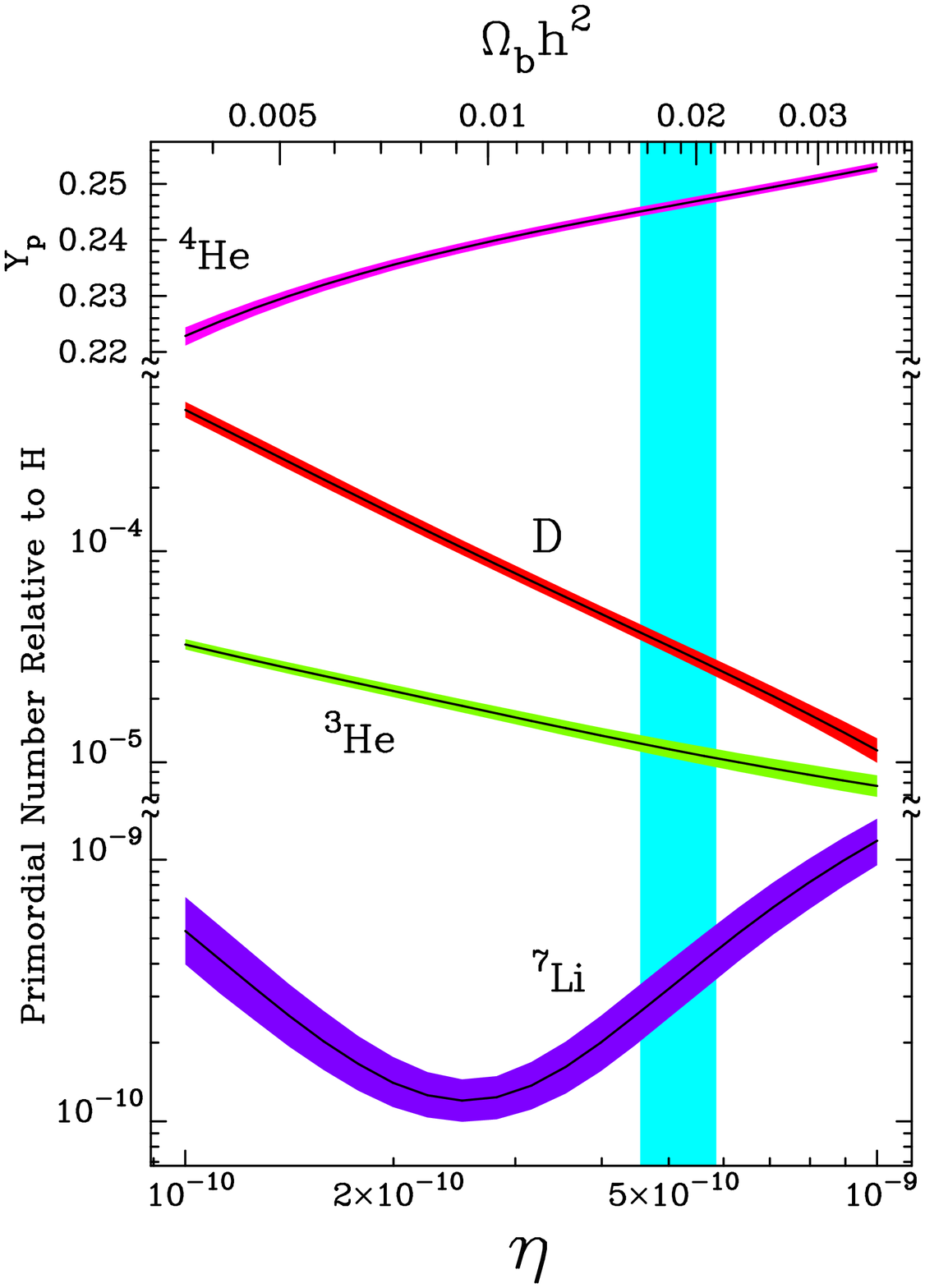,width=4.3in}}
\caption{Predicted abundances of $^4$He (mass fraction),
D, $^3$He, and $^7$Li (number relative to hydrogen) as a function
of the baryon density; widths of the curves indicate
``$2\sigma$'' theoretical uncertainty.  The dark band highlights
the determination of the baryon density based upon
the recent measurement of the primordial abundance of
deuterium (Burles \& Tytler, 1998a,b), $\Omega_Bh^2 = 0.019 \pm
0.0024$ (95\% cl); the baryon density is related to the baryon-to-photon
ratio, $\rho_B = 6.88\eta \times 10^{-22}\gcmm3$ (from Burles et al, 1999).
}
\label{fig:BBN}
\end{figure}

The accounting of matter and energy goes like this
(in units of the critical density for $H_0=65\kms\Mpc^{-1}$):
light neutrinos, between 0.3\%  and 15\%; stars and related
material, between 0.3\% and 0.6\%; baryons (total), $5\%\pm 0.5\%$;
matter (total), $40\%\pm 10\%$; and vacuum
energy (or something similar), $80\%\pm 20\%$; within the uncertainties,
a total equalling the critical density (see Fig.~\ref{fig:omega_sum}).

The recently measured primeval
deuterium abundance (Burles \& Tytler, 1998a,b)
and the theory of big-bang nucleosynthesis now
accurately pin down the baryon density (Schramm \& Turner, 1998;
Burles et al, 1999), $\Omega_B = (0.019\pm 0.0012)
h^{-2} \simeq 0.05$ (for $h=0.65$); see Fig.~\ref{fig:BBN}.
Using the cluster baryon
fraction, determined from x-ray measurements (Mohr et al, 1998;
Evrard, 1996)
and SZ measurements (Carlstrom, 1999), $f_B = M_{\rm baryon}/
M_{\rm TOT} = (0.07\pm 0.007)h^{-3/2}$,
and assuming that clusters provide
a fair sample of matter in the Universe, $\Omega_B/\Omega_M
= f_B$, it follows that $\Omega_M = (0.3\pm 0.05)h^{-1/2} \simeq 0.4\pm 0.1$.
Other direct measurements of the matter density are consistent
with this (see e.g., Turner, 1999).

That $\Omega_M \gg \Omega_B$ is strong, almost incontrovertible, evidence
for nonbaryonic dark matter; the leading particle candidates are axions,
neutralinos and neutrinos.  The recent evidence for neutrino
oscillations, based upon atmospheric neutrino data presented
by the SuperKamiokande Collaboration, indicates that neutrinos
contribute at least as much mass as bright stars; the failure
of the, top-down, hot dark matter scenario of structure formation
restricts the contribution of neutrinos to be less than about
15\% of the critical density (see e.g., Dodelson et al, 1996;
White, Frenk \& Davis, 1983).  Because relic axions and neutralinos
behave like cold dark matter (i.e., move slowly), they are the
prime particle dark-matter candidates.

\begin{figure}
\centerline{\psfig{figure=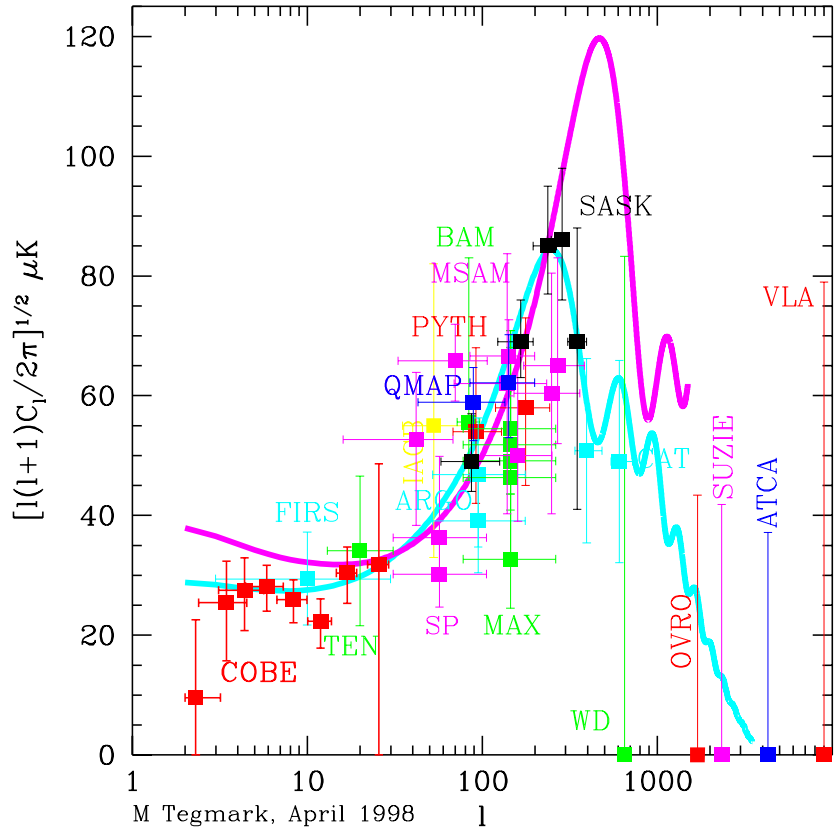,width=4.3in}}
\caption{Summary of current CBR anisotropy measurements, where
the temperature variation across the sky has been expanded
in spherical harmonics, $\delta T(\theta , \phi ) = \sum_i a_{lm}Y_{lm}$
and $C_l \equiv \langle |a_{lm}|^2\rangle$.  The curves
illustrate CDM models with $\Omega_0 = 1$ (lighter) and $\Omega_0 =0.3$
(darker).  Note the preference of the data for a flat Universe
(Figure courtesy of M. Tegmark).}
\label{fig:cbr_today}
\end{figure}

\begin{figure}
\centerline{\psfig{figure=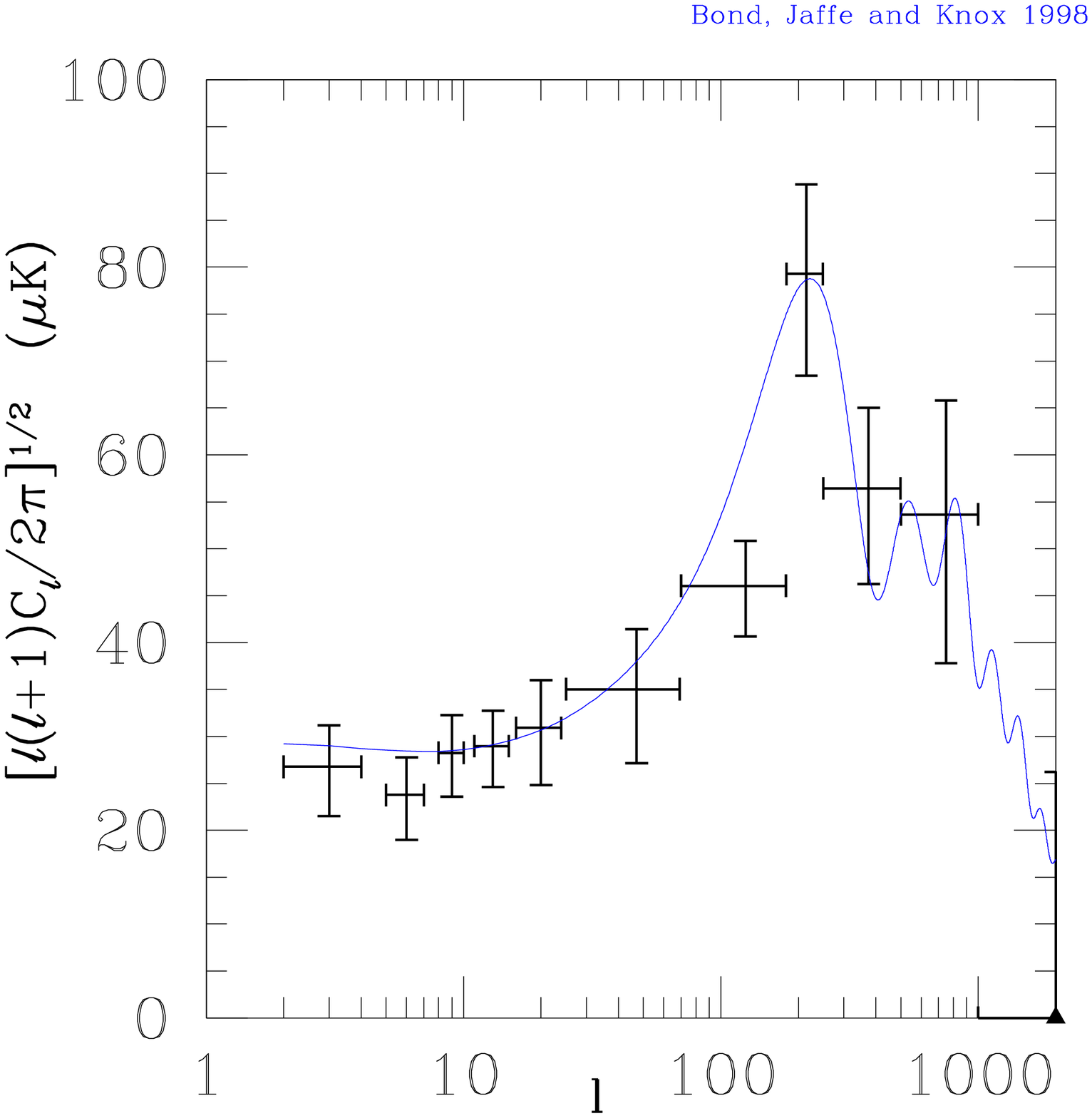,width=4.3in}}
\caption{The same data as in the previous figure, but
averaged and binned to reduce error bars and
visual confusion.  The theoretical
curve is for the $\Lambda$CDM model with $H_0=65\,{\rm km\,
s^{-1}\,Mpc^{-1}}$ and $\Omega_M =0.4$ (Figure courtesy of L. Knox).
}
\label{fig:cbr_knox}
\end{figure}

The position of the first acoustic peak in the angular power spectrum
of temperature fluctuations of the CBR
is a sensitive indicator of the curvature of the Universe:
$l_{\rm peak} \simeq 200/\sqrt{\Omega_0}$, where $R_{\rm curv}^2
= H_0^{-2}/|\Omega_0 -1|$.  CBR anisotropy measurements now span
multipole number $l=2$ to around $l=1000$
(see Figs.~\ref{fig:cbr_today} and \ref{fig:cbr_knox});
while the data do not yet speak definitively, it is clear that $\Omega_0 \sim
1$ is preferred.  Several experiments (Python V, Viper, MAT
and Boomerang) with new results around $l=30 -700$
should be reporting in soon.  Ultimately, the MAP
(launch in 2000) and Planck (launch in 2007)
satellites will cover $l=2$ to $l=3000$ with precision
limited essentially by sampling variance, and should determine
$\Omega_0$ to a precision of 1\% or better.

The same angular power spectrum that indicates $\Omega_0\sim 1$
also provides evidence that the primeval density perturbations
are of the kind predicted by inflation.  The inflation-produced
Gaussian curvature fluctuations lead to an angular
power spectrum with a series of well defined acoustic peaks.  While the
data at best define the first peak, they are good enough to
exclude many models where the density perturbations are isocurvature
(e.g., cosmic strings and textures):  in these models the predicted
spectrum is devoid of acoustic peaks (Allen et al, 1997; Pen et al, 1997).

The oldest approach to determining $\Omega_0$ is by measuring
the deceleration of the expansion.  Sandage's deceleration parameter,
$q_0 \equiv -({\ddot R}/R)_0/H_0^2 = {\Omega_0\over 2}[1+3p/\rho ]$,
depends upon both $\Omega_0$ and the equation of state, $p(\rho )$.
Because distant objects are seen at an earlier epoch, by measuring
the (luminosity) distance to objects as a function of
redshift the deceleration of the Universe can be determined.
(If the Universe is slowing down, distant
objects should be moving faster than predicted by Hubble's law,
$v_0=H_0d$.)  Accurate
distance measurements to some fifty supernovae of type Ia (SNe Ia)
carried out by two groups (Riess et al, 1998;
Perlmutter et al, 1998) indicate that the Universe is speeding
up, not slowing down (i.e., $q_0<0$).  The simplest explanation is
a cosmological constant, with $\Omega_\Lambda \sim 0.6$.  This
result makes the CBR determination of the total density ($\Omega_0 =1$)
and direct measures of the matter density ($\Omega_M\sim 0.4$)
consistent:  the ``missing energy'' exists in a smooth component that
cannot clump and thus is not found in clusters of galaxies.

\begin{figure}
\centerline{\psfig{figure=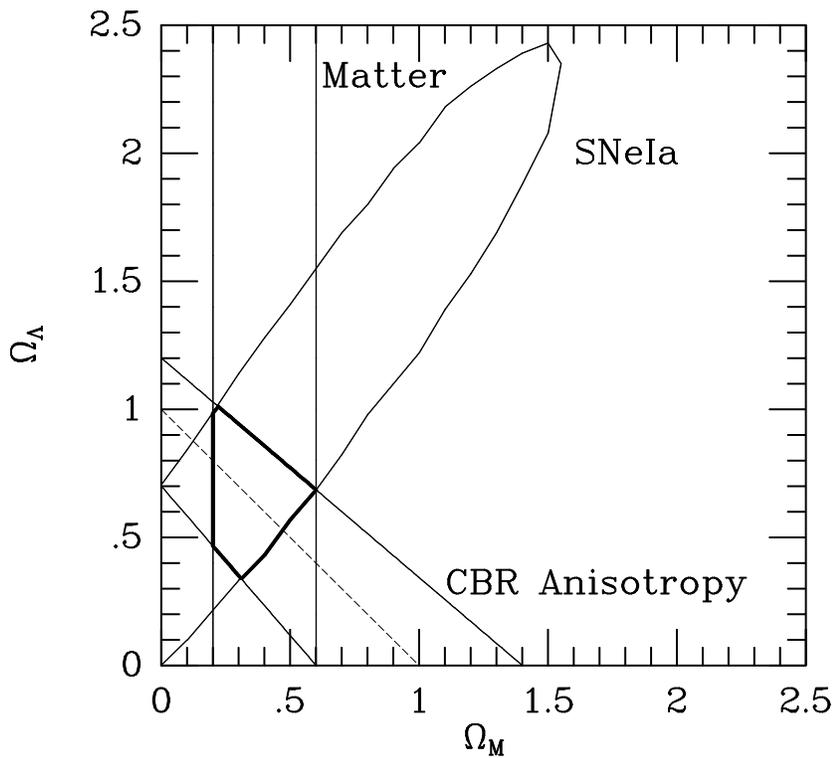,width=4.3in}}
\caption{Two-$\sigma$ constraints to $\Omega_M$ and $\Omega_\Lambda$
from CBR anisotropy, SNe Ia, and measurements of clustered matter.
Lines of constant $\Omega_0$ are diagonal, with a flat
Universe shown by the broken line.
The concordance region is shown in bold:  $\Omega_M\sim 1/3$,
$\Omega_\Lambda \sim 2/3$, and $\Omega_0 \sim 1$.
(Particle physicists who rotate the figure by $90^\circ$
will recognize the similarity to the convergence of the
gauge coupling constants.)
}
\label{fig:omegalambda}
\end{figure}

The concordance of the three measurements
that bear on the quantity and composition of matter and energy
in the Universe is illustrated in Fig.~\ref{fig:omegalambda}.
The SN Ia results are sensitive to the
acceleration (or deceleration) of the expansion
and constrain the combination ${4\over 3}\Omega_M -\Omega_\Lambda$.  (Note,
$q_0 = {1\over 2}\Omega_M - \Omega_\Lambda$; ${4\over 3}\Omega_M -
\Omega_\Lambda$ corresponds to the deceleration parameter
at redshift $z\sim 0.4$, the median redshift of these
samples).  The (approximately) orthogonal combination,
$\Omega_0 = \Omega_M + \Omega_\Lambda$
is constrained by CBR anisotropy.  Together, they define a concordance
region around $\Omega_0\sim 1$, $\Omega_M \sim 1/3$,
and $\Omega_\Lambda \sim 2/3$.  The constraint
to the matter density alone, $\Omega_M = 0.4\pm 0.1$,
provides a cross check, and it is consistent with these numbers.
Cosmic concordance!

While the evidence for inflation + cold dark matter
is not definitive and we should be cautious,
1998 could well mark a turning point in cosmology as important
as 1964.  Recall, after the discovery of the CBR
it took a decade or more to firmly
establish the cosmological origin of the CBR and
the hot big-bang cosmology as the standard cosmology.

\section{Inflation + Cold Dark Matter}

Inflation has revolutionized the way cosmologists view the Universe
and provides the current working hypothesis for extending the
standard cosmology to much earlier times.
It explains how a region of size much, much
greater than our Hubble volume could have become smooth and flat
without recourse to special initial conditions (Guth 1981), as well as
the origin of the density inhomogeneities
needed to seed structure (Hawking, 1982; Starobinsky, 1982;
Guth \& Pi, 1982; and Bardeen et al, 1983).
Inflation is based upon
well defined, albeit speculative physics -- the semi-classical evolution
of a weakly coupled scalar field -- and this physics may well
be connected to the unification of the particles and forces of Nature.

It would be nice if there were a standard model of inflation, but
there isn't.   What is important, is that almost all inflationary
models make three very testable predictions:  flat
Universe, nearly scale-invariant
spectrum of Gaussian density perturbations, and nearly scale-invariant
spectrum of gravitational waves.  These three predictions allow the
inflationary paradigm to be decisively tested.  While the gravitational
waves are an extremely important and challenging test, I will not
mention them again here (see e.g., Turner 1997a).

The tremendous expansion that occurs during inflation
is key to its beneficial effects and robust predictions:
A small, subhorizon-sized bit
of the Universe can grow large enough to encompass the entire
observable Universe and much more.  Because all that
we can see today was once so extraordinarily small, it began
flat and smooth.  This is unaffected
by the expansion since then and so the Hubble radius today is
much, much smaller than the curvature radius, implying $\Omega_0
=1$.  Lastly, the tremendous expansion stretches quantum fluctuations
on truly microscopic scales ($\la 10^{-23}\rcm$)
to astrophysical scales ($\gg$ millions of light years).

The curvature perturbations created by inflation
are characterized by two important features: 1) they are
almost scale-invariant, which refers to the fluctuations in
the gravitational potential being independent
of scale -- and not the density perturbations themselves;
2) because they arise from fluctuations in an essentially noninteracting
quantum field, their statistical properties are that of
a Gaussian random field.

Scale invariance specifies the shape of the spectrum of density
perturbations.  The normalization (overall amplitude) depends
upon the specific inflationary model (i.e., scalar-field potential).
Ignoring numerical factors for the moment, the overall amplitude
is specified by the fluctuation in the gravitational
potential, $\delta \phi\simeq (\delta \rho /\rho )_{\rm HOR} \sim
V^{3/2}/m_{\rm PL}^3 V^\prime$, which is also equal to
the amplitude of density perturbations when they cross the horizon.
To be consistent with the COBE measurement of CBR anisotropy
on the $10^\circ$ scale, $\delta \phi$ must be around $2\times 10^{-5}$.
Not only did COBE produce the first evidence for the existence of
the density perturbations that seeded all structure
(Smoot et al, 1992), but also,
for a theory like inflation that predicts
the shape of the spectrum of density perturbations,
it fixed the amplitude of density perturbations on all scales.
The COBE normalization began precision testing of inflation.

\section{Cold Dark Matter -- A Cosmological Necessity!}

\begin{figure}
\centerline{\psfig{figure=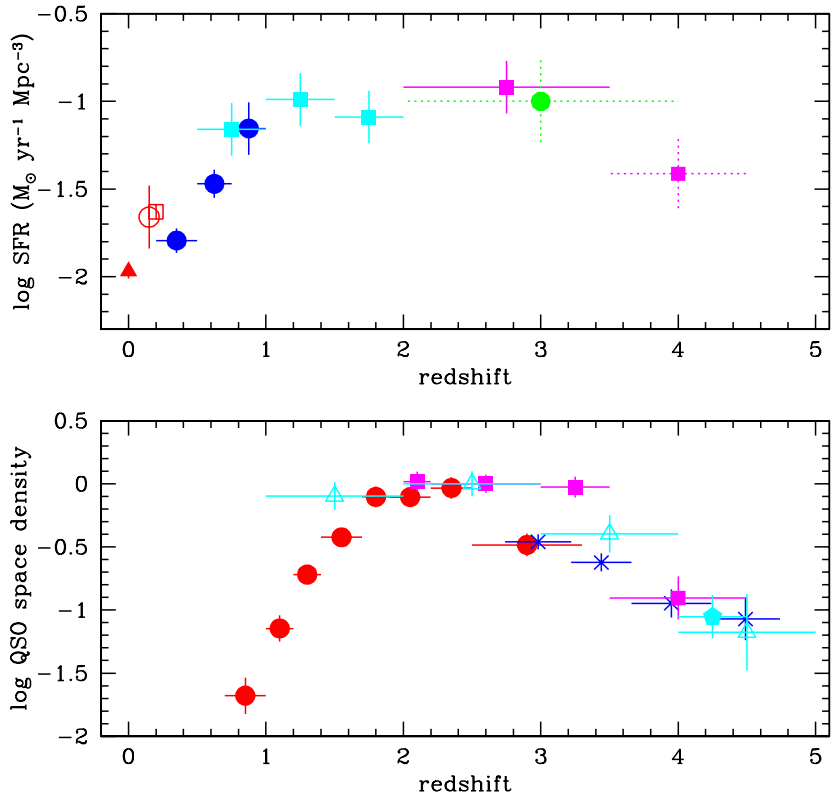,width=4.3in}}
\caption{The history of galaxy formation in the Universe,
as registered by star formation.  The top panel
shows the star formation rate vs redshift for galaxies, and
the bottom for QSOs (which are hosted by galaxies).
As can be seen in both panels, the epoch of galaxy formation
occurs at redshifts of a few, as predicted by CDM.
The points have been corrected for dust and the 
relative space density of QSOs (from Madau, 1999)
}
\label{fig:SFR}
\end{figure}

While we don't know what the cold dark matter consists of,
there is overwhelming evidence that it must be there.
(Generically, cold dark matter refers to
particles that comprise the bulk of the
matter density, move very slowly, interact feebly with ordinary
matter (baryons) and are not comprised of ordinary matter.)
The biggest surprise in cosmology that I can
imagine is the nonexistence of cold dark matter.
Here is a brief summary of the most compelling
evidence for cold dark matter:

\begin{itemize}

\item{}  For more than a decade there has been growing evidence that
the total amount of matter is significantly greater than what
baryons can account for.  Today,
the discrepancy is about a factor of eight:  $\Omega_M = 0.4
\pm 0.1$ and $\Omega_B = (0.02\pm 0.002)h^{-2} \simeq 0.05$.
Unless BBN is grossly misleading us
and/or determinations of the matter density are way
off, most of the matter must be nonbaryonic.
(The discovery of dark
energy does nothing to change this fact; it explains the discrepancy
between the matter density, $\Omega_M =0.4$, inferred from
matter that clusters, and the total density, $\Omega_0 = 1$,
inferred from CBR anisotropy.)

\item{}  We now know that galaxies formed at redshifts of order
2 to 4 (see Fig.~\ref{fig:SFR}), that clusters formed at redshifts of 1 or less and that
superclusters are forming today.  That is, structure formed from the
bottom up, as predicted if the nonbaryonic matter is cold.  (Hot
dark matter leads to a top-down sequence of structure formation;
see, White, Frenk, and Davis, 1983.)

\item{}  The cold dark matter model of structure formation is
consistent with an enormous body of data:  CBR anisotropy, large-scale
structure, abundance of clusters, the clustering of galaxies
and clusters, the evolution of clusters and galaxies and
their clustering, the structure of the Lyman-$\alpha$
forest, and a host of other data.

\item{} The only plausible candidate for the bulk of the dark
matter in the halo of own galaxy is cold dark matter particles.
The last-hope baryonic candidate, dark stars or MACHOs, can account for only
about half the mass of the halo and probably much less.  This follows
from the fact the microlensing rates toward the Magellanic
Clouds, which are about 30\% of that expected if the halo
were comprised entirely of MACHOs, and the growing evidence
that the Magellanic lenses are in the clouds themselves or
other nonhalo components of the Galaxy.

\end{itemize}

The two leading particle candidates for cold dark matter
are the axion and the neutralino.  Both are well motivated
by fundamental physics concerns and both have a predicted
relic abundance that is comparable to the critical density.
There are other ``dark-horse'' candidates including primordial
black holes and superheavy relic particles, which should
not be forgotten (Kolb, 1999).  As far as cosmological infrastructure
goes, they would be every bit as good as axions and neutralinos.

\section{Neutrinos by the Numbers}

Cosmic neutrinos are almost abundant as CBR photons:
$n_{\nu\bar\nu} = {3\over 11}n_\gamma$ (per species)
$\simeq 112\cmm3$.  Cosmologists are confident
of their relic abundance (at least within the standard
model of particle physics) because they were in thermal
equilibrium until the Universe was a second old; thereafter,
their weak interactions were too ``weak'' to keep them in thermal
equilibrium and their temperature decreased as $R^{-1}$.
Shortly after neutrinos ``decoupled''
electrons and positrons annihilated, raising the photon
temperature slightly so that $T_\nu/T_\gamma = (4/11)^{1/3}$.
Because the yields of big-bang nucleosynthesis are so
sensitive to the phase-space distribution of neutrinos,
the success of BBN is also a confirmation of the
standard cosmic history of neutrinos.

\begin{figure}
\centerline{\psfig{figure=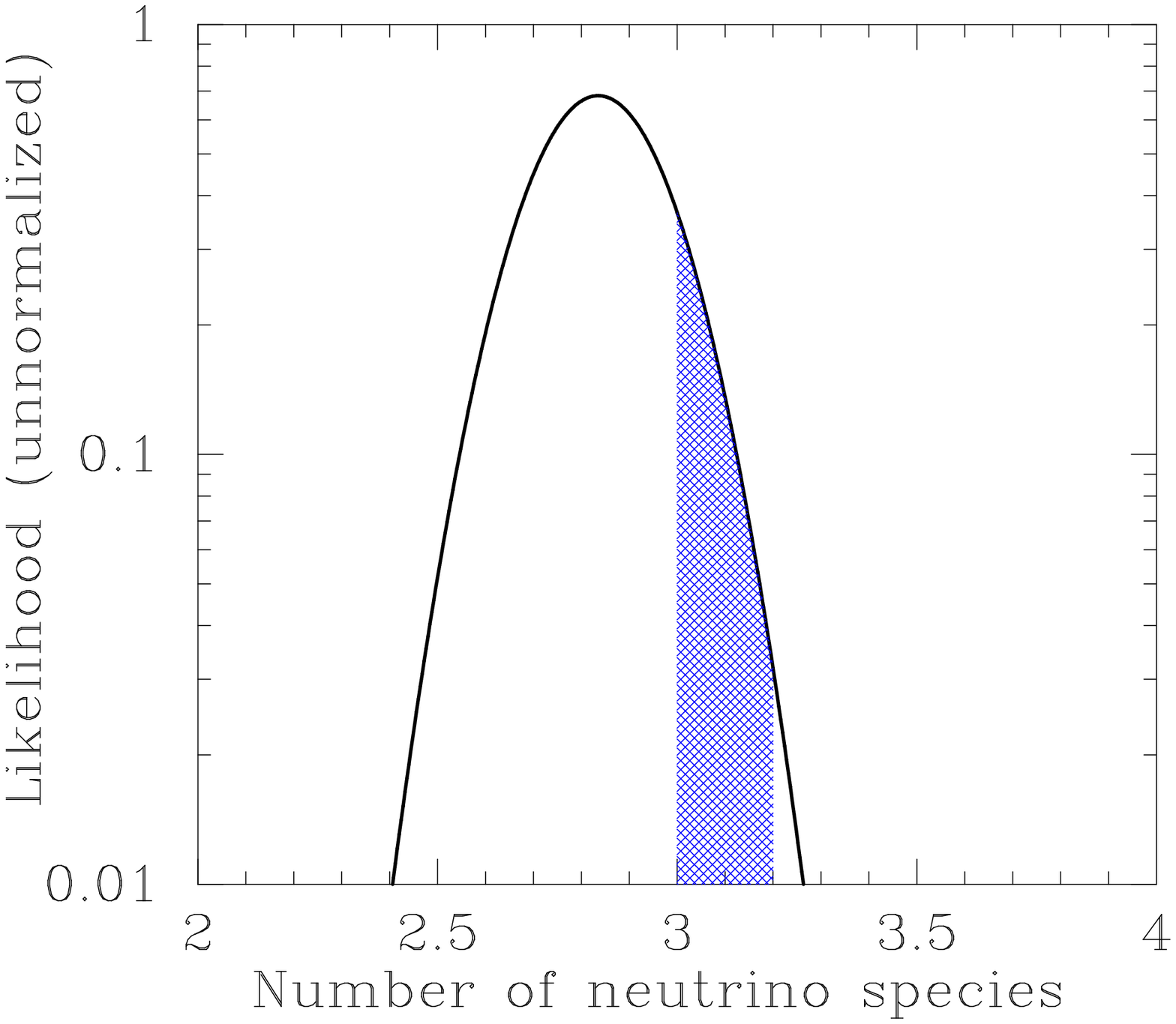,width=4.3in}}
\caption{Likelihood function for big-bang production
of $^4$He as a function of $N_\nu$.  The BBN yield
of $^4$He (mass fraction $Y_P$) increases with
increasing $N_\nu$; for this analysis the deuterium-determined
baryon density was assumed and $Y_P=0.244\pm 0.002$.
The 95\% confidence region is $N_\nu = 2.84\pm 0.3$;
the 95\% cl upper limit (with the prior $N_\nu \ge 3$) is
$N_\nu < 3.2$; shaded area indicates 95\% confidence
region for $N_\nu > 3$ prior (Figure courtesy of Scott Burles).
}
\label{fig:like}
\end{figure}

The sensitivity of BBN to neutrinos allowed Steigman, Schramm
and Gunn (1977) to use the yield of $^4$He to constraint
the number of light neutrino species.  Their original limit,
$N_\nu < 7$, bettered the laboratory limit at the
time by almost a factor of 1000.  A recent analysis
(Burles et al, 1999) finds $N_\nu = 2.84\pm 0.3$ (95\% cl;
see Fig.~\ref{fig:like}), not quite as good as the LEP
determination of $3.07\pm 0.24$,
but still very impressive.  Since we are convinced that there
are just three standard neutrinos, both the LEP and BBN
determinations are now used to search for the existence of
new particles; in the case of BBN, light (mass less than
about 1\,MeV) species with sufficiently potent interactions
to be present in significant numbers around the time of BBN.
The current BBN limit, with the prior $N_\nu \ge 3$, is:
$N_\nu < 3.2$ (95\% cl).

Neutrinos were the first candidate for nonbaryonic dark matter
(motivated by since refuted evidence for
a 30\,eV electron neutrino mass in 1978), and this led to
the hot dark matter theory of structure formation.  While
many of its features are qualitatively correct, e.g., the existence
of voids, walls and sheets, it predicted ``top down'' formation of
structure (White, Frenk \& Davis, 1983) and it is now very
clear that structure formed from the ``bottom up.''

Because they are known to exist and are so abundant, neutrinos may
well make up a significant part of the mass budget and be
an interesting cosmic spice.  Here are the numbers:
\begin{eqnarray}
\Omega_\nu & = & {m_\nu  \over 90h^2\eV}
        \simeq {m_\nu \over 40\eV} \nonumber\\
\Omega_\nu/\Omega_B & = & {m_\nu \over 1.7\eV} \nonumber \\
\Omega_\nu /\Omega_* & = & {m_\nu \over 0.3h\eV}
        \simeq {m_\nu \over 0.2\eV}\nonumber
\end{eqnarray}
The SuperKamiokande data, which indicate a neutrino mass-squared
difference of around $10^{-2}\eV^2$, put the neutrino contribution
to the cosmic mass budget at an amount at least comparable to that
of bright stars.  WOW!  If the $0.1\eV$ mass corresponds to the
lightest neutrino species or if the mass difference squared
arises from nearly degenerate neutrino species, the total could
even greater -- $\sum_i m_{\nu_i} \simeq 1.7\eV$ would make neutrinos as
important as baryons.

A neutrino mass of a few tenths of an eV is already very interesting
from the point of view of large-scale structure formation.  Hu et al
(1998) have shown a neutrino species of this mass can have a potentially
detectable influence on large-scale structure, one which could well
be detectable with Sloan Digital Sky Survey data.  (When CBR
anisotropy data are folded in as well, the detection mass-limit might
well be even lower.)  At the other extreme, the effect on the formation
of large-scale structure is so profound (and bad), that $\Omega_\nu > 0.15$
($m_\nu \sim 4\eV$) can already be ruled out (Dodelson et al, 1996).

\begin{figure}
\centerline{\psfig{figure=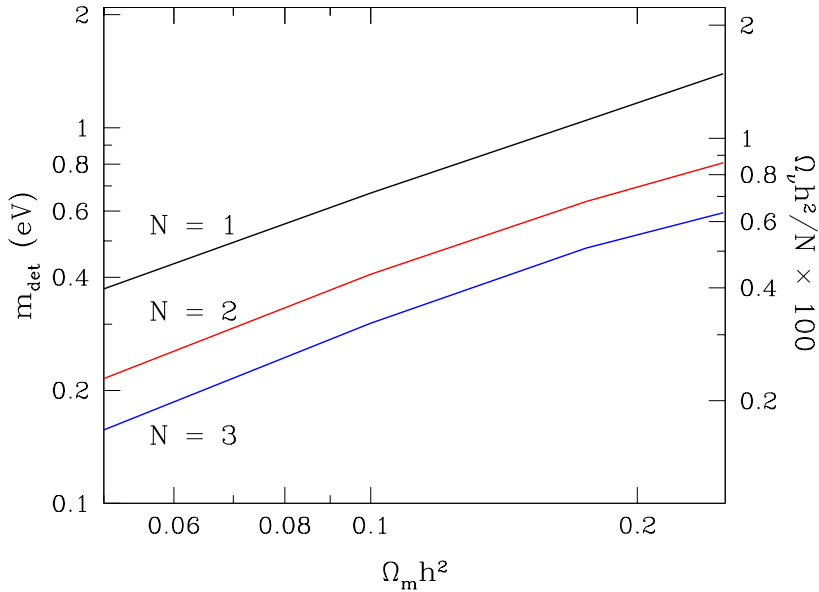,width=4.3in}}
\caption{Projected ``$2-\sigma$'' neutrino-mass detection limit
using Sloan Digital Sky Survey large-scale structure
measurements.  $N$ is the number of neutrino
species with the mass indicated and $\Omega_Mh^2\sim
0.15$ is the total mass density (from Hu et al, 1998).
}
\label{fig:nu_det}
\end{figure}

Finally, in the context of physics beyond the standard model,
cosmology still has much to tell us about neutrinos.  The properties
of massive, decaying neutrinos can be severely constrained by
BBN (Dodelson et al, 1994) and the CBR (Lopez et al, 1998).
Further, a massive tau neutrino that decays a few
seconds after the bang or later and produces relativistic particles,
can change the balance of matter and radiation, leading to
an interesting variant of cold dark matter called
$\tau$CDM (see below, and Dodelson et al, 1996).

\section{Inflation + CDM in the Era of Precision Cosmology}

As we look forward to the abundance (avalanche!) of high-quality observations
that will test inflation + CDM, we have to make sure the predictions
of the theory match the precision of the data.  In
so doing, CDM + inflation becomes a ten (or more) parameter
theory.  For astrophysicists, and especially cosmologists,
this is daunting, as it may seem that a ten-parameter
theory can be made to fit any set of observations.  This is
not the case when one has the quality and quantity of data
that will be coming.  The standard model of particle physics offers
an excellent example:  it is
a nineteen-parameter theory and because of the high-quality of
data from experiments at Fermilab's Tevatron, SLAC's SLC,
CERN's LEP and other facilities it has been rigorously tested
and the parameters measured to a precision of better than 1\%
in some cases.  My worry as an inflationist is not that many different
sets of parameters will fit the upcoming data, but rather that
no set will!

In fact, the ten parameters of CDM + inflation
are an opportunity rather than a curse:  Because the parameters
depend upon the underlying inflationary model and fundamental
aspects of the Universe, we have the very real possibility of learning
much about the Universe and inflation.  The ten parameters
can be organized into two groups:  cosmological
and dark-matter (Dodelson et al, 1996).

\smallskip
\centerline{\it Cosmological Parameters}
\begin{enumerate}

\item $h$, the Hubble constant in units of $100\kms\Mpc^{-1}$.

\item $\Omega_Bh^2$, the baryon density.  BBN implies:
$\Omega_Bh^2 = 0.019\pm 0.0024$ (95\% cl).

\item $n$, the power-law index of the scalar density perturbations.
CBR measurements indicate $n=1.1\pm 0.2$; $n=1$ corresponds to
scale-invariant density perturbations.  Most models predict $n\approx 0.90
-0.98$; the range of predictions runs from $0.7$ to $1.2$
(Lyth \& Riotto, 1996).

\item $dn/d\ln k$, ``running'' of the scalar index with comoving scale
($k=$ wavenumber).  Most models predict a value of
${\cal O}(\pm 10^{-3})$ or smaller (Kosowsky \& Turner, 1995).

\item $S$, the overall amplitude squared of density perturbations,
quantified by their contribution to the variance of the
CBR quadrupole anisotropy.

\item $T$, the overall amplitude squared of gravity waves,
quantified by their contribution to the variance of the
CBR quadrupole anisotropy.  Note, the COBE normalization determines
$T+S$.

\item $n_T$, the power-law index of the gravity wave spectrum.
Scale-invariance corresponds to $n_T=0$; for inflation, $n_T$
is given by $-{1\over 7}{T\over S}$.

\end{enumerate}

\centerline{\it Dark-matter Parameters}

\begin{enumerate}

\item $\Omega_\nu$, the fraction of critical density in neutrinos.
While the hot dark matter theory of structure
formation is not viable, it is possible that a small fraction
($\Omega_\nu < 0.15$) of the matter density exists in the form of neutrinos.

\item $\Omega_X$, the fraction of critical density in a smooth component
of unknown composition and negative pressure ($w_X \la -0.3$).  The
SN Ia results and CBR anisotropy provide strong evidence for such a component,
with the simplest example being a cosmological constant ($w_X = -1$).

\item $g_*$, the quantity that counts the number of ultra-relativistic
degrees of freedom around the time of matter-radiation
equality.  In the standard cosmology/standard
model of particle physics $g_* = 3.3626$ (photons in the
CBR + 3 massless neutrino species).  The amount of radiation controls when
the Universe became matter dominated and thus affects the present
spectrum of matter inhomogeneity.

\end{enumerate}

As mentioned, the parameters involving density and gravity-wave
perturbations depend directly upon the inflationary potential.
In particular, they can be expressed in terms of the potential
and its first three derivatives:

\begin{eqnarray}
S  & \equiv & {5\langle |a_{2m}|^2\rangle \over 4\pi} \simeq
         2.2\,{V_*/m_{\rm Pl}^4 \over (m_{\rm Pl} V_*^\prime /V_*)^2}
         \nonumber \\
n -1 & = & -{1\over 8\pi}\left({\mpl V^\prime_* \over V_*} \right)^2
        + {\mpl \over 4\pi}\left( {\mpl V_*^{\prime}\over V_*} \right)^\prime
        \nonumber \\
{dn\over d\ln k} & = &
-{1\over 32\pi^2}\left({{m_{\rm Pl}}^3V_*^{\prime\prime\prime}\over
        V_*}\right)
        \left({{m_{\rm Pl}} V_*^\prime\over V_*}\right) \nonumber\\
     & & \hspace*{1cm} + {1\over 8\pi^2}
        \left({{m_{\rm Pl}}^2V_*^{\prime\prime}\over V_*}\right)
        \left({{m_{\rm Pl}} V_*^\prime\over V_*}\right)^2 \nonumber\\
     & &- {3\over 32\pi^2}\left({m_{\rm Pl}} {V_*^\prime\over V_*}\right)^4
     \nonumber \\
T  & \equiv & {5\langle |a_{2m}|^2\rangle \over 4\pi} =
        0.61 (V_*/m_{\rm Pl}^4) \nonumber\\
n_T & = & -{1\over 8\pi} \left( {\mpl V^\prime_* \over V_*} \right)^2
        \nonumber
\end{eqnarray}
where $V(\phi )$ is the inflationary potential, prime denotes $d/d\phi$,
and $V_*$ is the value of the scalar potential when the present horizon
scale crossed outside the horizon during inflation.

If one can measure $S$, $T$, and $(n-1)$,
one can recover the value of the potential
and its first two derivatives (see e.g.,
Turner 1993; Lidsey et al, 1997)
\begin{eqnarray}
V_* & = & 1.65 T\, {m_{\rm Pl}}^4  , \\
V_*^\prime & = & \pm \sqrt{{8\pi \over 7}\,{T\over S}}\, V_*/{m_{\rm Pl}} , \\
V_*^{\prime\prime} & = & 4\pi \left[ (n-1) + {3\over 7} {T\over S} \right]\,
V_* /{m_{\rm Pl}}^2 ,
\end{eqnarray}
where the sign of $V^\prime$ is indeterminate (under the
redefinition $\phi \leftrightarrow
-\phi$ the sign changes).  If, in addition, the gravity-wave spectral index
can also be measured the consistency relation, $T/S = -7 n_T$,
can be used to test inflation.

Bunn \& White (1997) have used the
COBE four-year dataset to determine $S$ as a function of
$T/S$ and $n-1$; they find
\begin{eqnarray}
{V_*/m_{\rm Pl}^4 \over (m_{\rm Pl} V_*^\prime /V_*)^2}
        & = & {S\over 2.2} =
        (1.7\pm 0.2)\times 10^{-11} \nonumber\\
        & & \times {\exp [-2.02(n-1)] \over
        \sqrt{1+{2\over 3}{T\over S}}}
\end{eqnarray}
{}From which it follows that
\begin{equation}
V_* < 6\times 10^{-11} \mpl^4,
\end{equation}
equivalently, $V_*^{1/4} < 3.4\times 10^{16}\GeV$.
This indicates that inflation must involve
energies much smaller than the Planck scale.
(To be more precise, inflation could have begun at a much higher
energy scale, but the portion of inflation relevant for us, i.e.,
the last 60 or so e-folds, occurred at an energy scale much smaller
than the Planck energy.)

Finally, it should be noted that the `tensor tilt,'
deviation of $n_T$ from 0, and the `scalar
tilt,' deviation of $n-1$ from zero, are not in general equal;
they differ by the rate of change of the steepness.
The tensor tilt and the ratio $T/S$ are related:
$n_T = -{1\over 7}{T\over S}$, which provides a
consistency test of inflation.

\subsection{Present status of Inflation + CDM}

\begin{figure}
\centerline{\psfig{figure=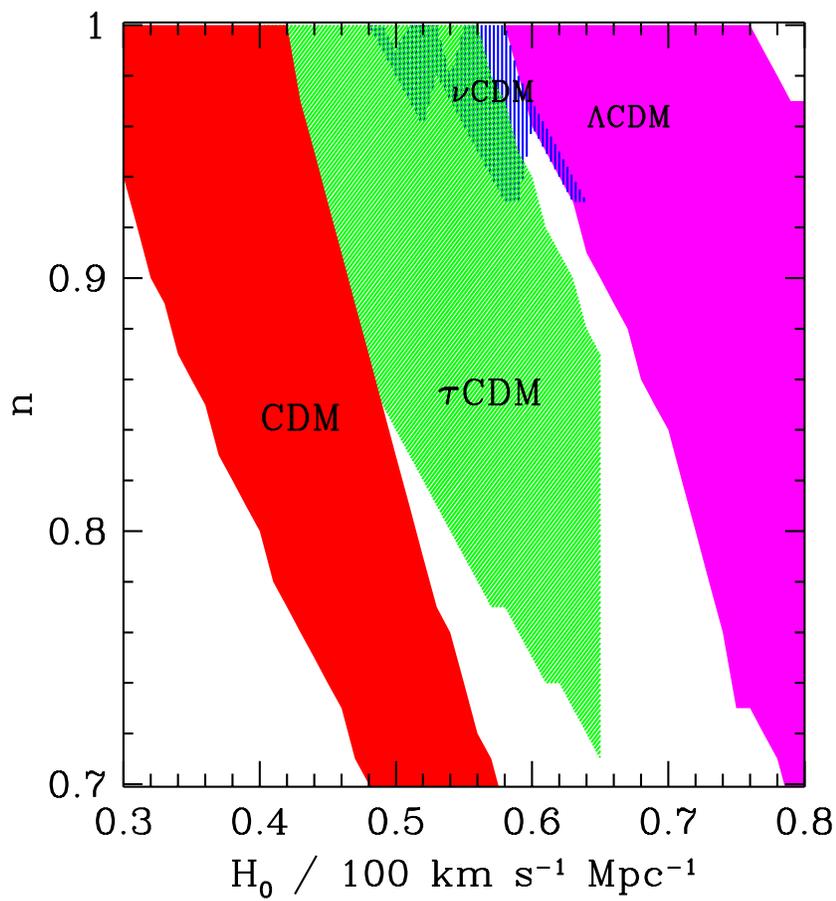,width=4.3in}}
\caption{Summary of viable CDM models, based upon
CBR anisotropy and determinations of the present
power spectrum of inhomogeneity (Dodelson et al, 1996).}
\label{fig:cdm_sum}
\end{figure}

A useful way to organize the different CDM models is by their
dark-matter content; within each CDM family, the cosmological
parameters vary.  One classification is (Dodelson et al, 1996):

\begin{enumerate}

\item sCDM (for simple):  Only CDM and baryons; no additional
radiation ($g_*=3.36$).  The original standard CDM is a member
of this family ($h=0.50$, $n=1.00$, $\Omega_B=0.05$), but is
now ruled out (see Fig.~\ref{fig:cdm_sum}).

\item $\tau$CDM:  This model has
extra radiation, e.g., produced by the decay of an unstable
massive tau neutrino (hence the name); here we take $g_* = 7.45$.

\item $\nu$CDM (for neutrinos):  This model has a dash of hot
dark matter; here we take $\Omega_\nu = 0.2$ (about 5\,eV
worth of neutrinos).

\item $\Lambda$CDM (for cosmological constant):  This model has
a smooth component in the form of a cosmological constant; here
we take $\Omega_\Lambda = 0.6$.

\end{enumerate}

\begin{figure}
\centerline{\psfig{figure=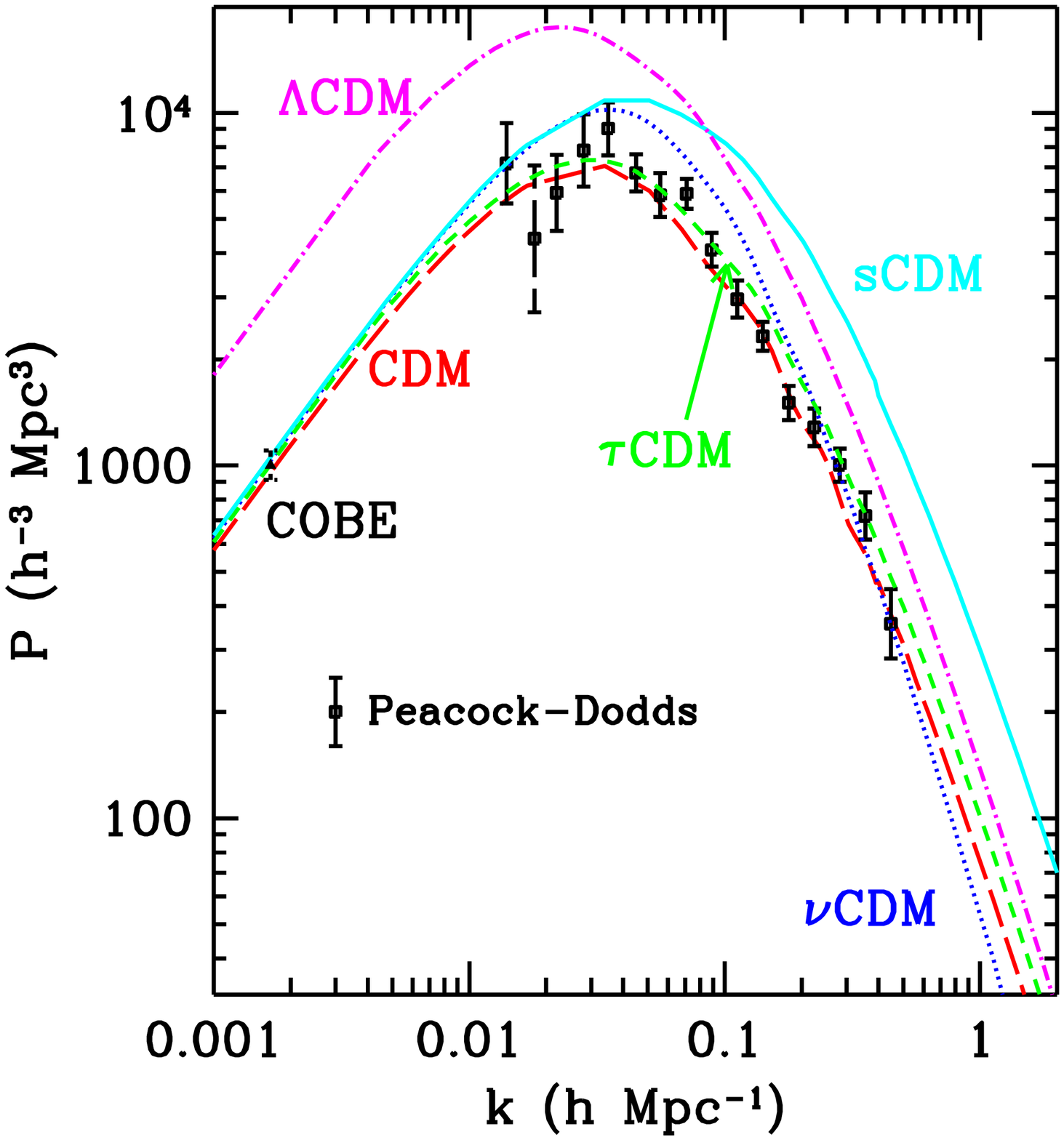,width=4.3in}}
\caption{The power spectrum of fluctuations today, as traced by
bright galaxies (light), as derived from redshift surveys assuming
light traces mass (Peacock and Dodds, 1994).  The curves correspond to the 
predictions of various cold dark matter models.   The relationship
between the power spectrum and CMB anisotropy
in a $\Lambda$CDM model is different, and in fact, the $\Lambda$CDM
model shown is COBE normalized.
}
\label{fig:pd}
\end{figure}

Figure \ref{fig:cdm_sum} summarizes the viability of
these different CDM models,
based upon CBR measurements and current determinations of
the present power spectrum of inhomogeneity derived from
redshift surveys (see Fig.~\ref{fig:pd}).
sCDM is only viable for low values of the
Hubble constant (less than $55\kms\Mpc^{-1}$) and/or
significant tilt (deviation from scale invariance); the region
of viability for $\tau$CDM is similar to sCDM, but shifted
to larger values of the Hubble constant (as large as
$65\kms\Mpc^{-1}$).  $\nu$CDM has an island of viability
around $H_0\sim 60\kms\Mpc^{-1}$ and $n\sim 0.95$.  $\Lambda$CDM
can tolerate the largest values of the Hubble constant.

\subsection{The best fit Universe!}

\begin{figure}
\centerline{\psfig{figure=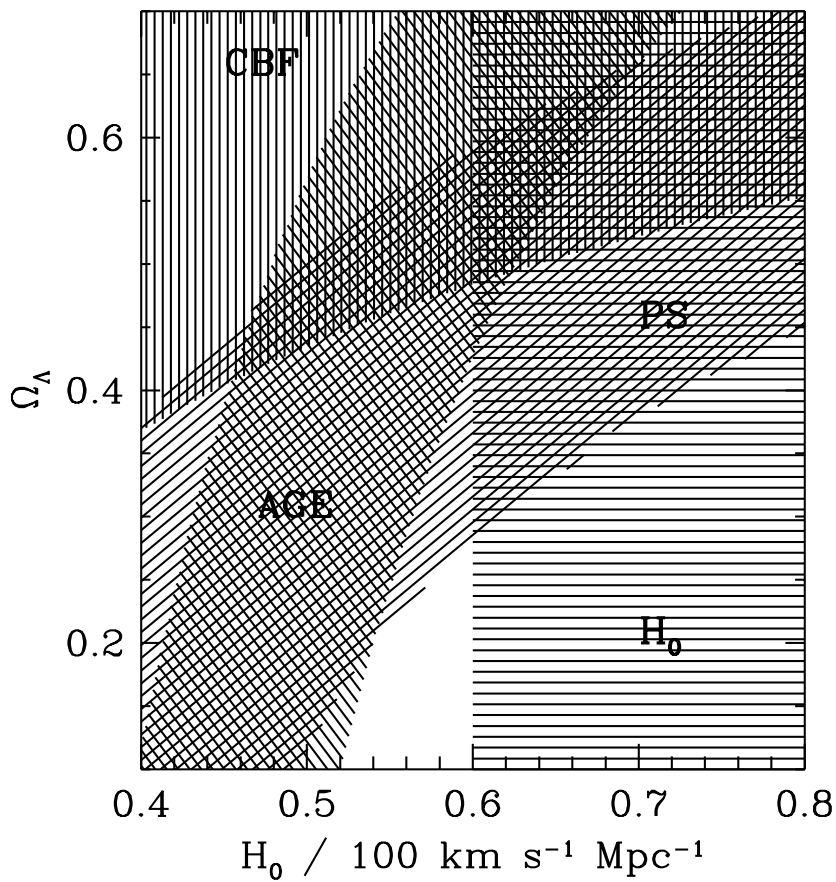,width=4.3in}}
\caption{Constraints used to determine the best-fit CDM model:
PS = large-scale structure + CBR anisotropy; AGE = age of the
Universe; CBF = cluster-baryon fraction; and $H_0$= Hubble
constant measurements.  The best-fit model, indicated by
the darkest region, has $h\simeq 0.60-0.65$ and $\Omega_\Lambda
\simeq 0.55 - 0.65$.}
\label{fig:best_fit}
\end{figure}

Considering other relevant data too -- e.g.,
age of the Universe, determinations of $\Omega_M$,
measurements of the Hubble constant, and limits to
$\Omega_\Lambda$ -- $\Lambda$CDM emerges as the
`best-fit CDM model' (Krauss \& Turner, 1995;
Ostriker \& Steinhardt, 1995; Liddle et al, 1996;
Turner, 1997b); see Fig.~\ref{fig:best_fit}.
Moreover, its `smoking-gun signature,' accelerated expansion,
has apparently been confirmed (Riess et al, 1998;
Perlmutter et al, 1998) and it provides an excellent
fit to the current CBR anisotropy data (see Fig.~\ref{fig:cbr_knox}).

Despite my general enthusiasm, I would caution that it is premature
to conclude that $\Lambda$CDM is anything but the model
to take aim at.  Further, it should be noted that the
SN Ia data do not yet discriminate between a cosmological
constant and something else with large, negative pressure
(e.g., rolling scalar field or frustrated topological
defects).

\section{Checklist for the Next Decade}

As I have been careful to stress the basic tenets of inflation
+ CDM have not yet been confirmed definitively.
However, a flood of high-quality cosmological data is
coming, and could make the case in the next decade.
Here are some of the important aspects of inflation
+ CDM that will be addressed by these data:

\begin{itemize}

\item Map of the Universe at 300,000 yrs.  COBE mapped
the CMB with an angular resolution of around $10^\circ$;
two new satellite missions, NASA's MAP (launch 2000)
and ESA's Planck Surveyor (launch 2007), will map the
CMB with 100 times better resolution ($0.1^\circ$). From
these maps of the Universe as it existed at a
simpler time, long before the first stars and
galaxies, will come a gold mine of information:
Among other things, a definitive measurement of $\Omega_0$;
a determination of the Hubble constant to a precision of
better than 5\%;
a characterization of the primeval lumpiness; and possible
detection of the relic gravity waves from inflation.
The precision maps of the CMB that will be made are crucial
to establishing inflation + cold dark matter.

\item Map of the Universe today.  Our knowledge of the structure
of the Universe is based upon maps constructed from the positions
of some 30,000 galaxies in our own backyard.  The Sloan Digital
Sky Survey will
produce a map of a representative portion of the Universe,
based upon the positions of a million galaxies.
The Anglo-Australian 2-degree Field survey will determine the position of
several hundred thousand galaxies.  These surveys will
define precisely the large-scale structure that exists today,
answering questions such as, ``What are the largest structures
that exist?''  Used together with
the CMB maps, this will definitively test the CDM
theory of structure formation, and much more.

\item Present expansion rate $H_0$.  Direct measurements
of the expansion rate using standard candles, gravitational time
delay, SZ imaging and the CMB maps will pin down the elusive
Hubble constant once and for all.  It is the fundamental parameter
that sets the size -- in time and space -- of the observable Universe.
Its value is critical to testing the self consistency of CDM.

\item Cold dark matter.  A key element of theory is the
cold dark matter particles that hold the Universe together;
until we actually
detect cold dark matter particles, it will be difficult to argue
that cosmology is solved.
Experiments designed to detect the dark matter that
holds are own galaxy together are now operating with sufficient
sensitivity to detect both neutralinos and axions
(see e.g., Sadoulet, 1999; or van Bibber et al, 1998).
In addition, experiments at particle accelerators (Fermilab
and CERN) will be hunting for the neutralino and its other
supersymmetric cousins.

\item Nature of the dark energy.  If the Universe is indeed accelerating,
then most of the critical density exists in the form of dark energy.  This
component is poorly understood.  Vacuum energy is only the
simplest possibly for the smooth dark component; there are
other possibilities:  frustrated topological defects (Vilenkin, 1984;
Pen \& Spergel, 1998)
or a rolling scalar field (see e.g., Ratra \& Peebles, 1998;
Frieman et al, 1995; Coble et al, 1997; Caldwell et al, 1998;
Turner \& White, 1997).   Independent evidence for the existence
of this dark energy, e.g., by CMB anisotropy, the SDSS and 2dF
surveys, or gravitational
lensing, is crucial for verifying the accounting of matter and energy
in the Universe I have advocated.  Additional measurements of SNe Ia could
help shed light on the precise nature of the dark energy.  The dark
energy problem is not only of great importance for cosmology, but
for fundamental physics as well.  Whether it is vacuum energy or
quintessence, it is a puzzle for fundamental physics and possibly
a clue about the unification of the forces and particles.

\end{itemize}

\section{New Questions; Some Surprises?}

Will cosmologists look back on 1998 as a year that rivals
1964 in importance?  I think it is quite possible.  In
any case, the flood of data that is coming will make the
next twenty years in cosmology very exciting.  It could be
that my younger theoretical colleagues will get their wish --
inflation + cold dark matter is falsified and it's back
to the drawing board.  Or, it may be that it is roughly correct,
but the real story is richer and even more interesting.
This happened in particle
physics.  The quark model of the 1960s was based upon
an approximate global $SU(3)$ flavor symmetry,
which shed no light on the dynamics
of how quarks are held together.  The standard model of particle
physics that emerged and which provides a fundamental description of
physics at energies less than a few hundred GeV,
is based upon the $SU(3)$ color gauge theory of quarks and gluons (QCD)
and the $SU(2)\otimes U(1)$ gauge theory of the electroweak interactions.
The difference between global and local $SU(3)$ symmetry was profound.

Even if inflation + cold dark matter does pass the series of stringent
tests that will confront it in the next decade, there will
be questions to address and issues to work out.  Exactly how does
inflation work and fit into the scheme of the unification of the
forces and particles?  Does the quantum gravity era of cosmology,
which occurs before inflation, leave a detectable imprint on the
Universe?  What is the topology of the Universe and are there
additional spatial dimensions?  Precisely how did the excess of matter
over antimatter develop?  What happened before inflation?  What
does inflation + CDM teach us about the unification
of the forces and particles of Nature?  Last, but certainly not least,
we must detect and identify the cold dark matter particles.

We live in exciting times!

\end{document}